\def\statusstringone{Proc.\ 2016 International Conference on 
                     Signal Processing and Communications, 
                     Bangalore, India.}

\def\versionstringone{\texttt{\statusstringone}}
\def\versionstringtwo{}
\def\versionstringthr{}

\documentclass[conference]{IEEEtran}
\IEEEoverridecommandlockouts
\ifCLASSINFOpdf
\else
\fi
\usepackage{amssymb} 
\usepackage{amsmath} 
\usepackage{color}

\usepackage{theorem} 

\usepackage{latexsym} 
\usepackage{epsfig} 
\usepackage{psfrag} 
\usepackage{bm, bbm} 
\usepackage{xspace} 
\usepackage{graphicx} 
\usepackage{cite} 
\usepackage{url} 
\usepackage{array}
\usepackage{subfigure,afterpage, wrapfig, pifont}
\usepackage{watermark}
\renewcommand{\baselinestretch}{1}
\renewcommand{\mathbf}[1]{{\bm{#1}}}

\newcommand{\ignore}[1]{}

\newcommand{\eg}{\emph{e.g.}}
\newcommand{\etal}{\emph{et al.}}
\newcommand{\ie}{\emph{i.e.}}
\newcommand{\etc}{\emph{etc.}}

\renewcommand{\leq}{\leqslant}
\renewcommand{\geq}{\geqslant}

\newcommand{\R}{\mathbb{R}}

\newcommand{\matr}[1]{\mathbf{#1}}
\newcommand{\vect}[1]{\mathbf{#1}}

\newcommand{\set}[1]{\mathcal{#1}}

\newcommand{\graph}[1]{\mathsf{#1}}

\newcommand{\defeq}{\triangleq}

\newcommand{\setA}{\set{A}}

\newcommand{\setE}{\set{E}}

\newcommand{\va}{\vect{a}}

\newcommand{\hva}{\vect{\hat a}}

\newcommand{\vx}{\vect{x}}

\newcommand{\tvx}{\vect{\tilde x}}

\newcommand{\tr}{\mathsf{T}}

\newcommand{\graphN}{\graph{N}}

\newcommand{\ZBethe}{Z_{\mathrm{B}}}
\newcommand{\ZBetheM}[1]{Z_{\mathrm{B}, #1}}

\newcommand{\cover}[1]{\tilde{#1}}
\newcommand{\cset}[1]{\cover{\set{#1}}}
\newcommand{\cgraph}[1]{\cover{\graph{#1}}}

\newcommand{\vxsampletot}{\tvx}

\newcommand{\Qxoldsample}[1]{Q_{\vx_{\textrm{sample}}}} 

\newcommand{\Qxsample}[1]{Q_{\vxsampletot}}

\newcommand{\hzero}{\hat 0}
\newcommand{\hone}{\hat 1}
\newcommand{\hcf}{\hat{f}}

\newcommand{\mdcnfg}[1]{\cover{\graph{#1}}_{\mathrm{MDC}}}
\newcommand{\tmdcnfg}[1]{\hat{\graph{#1}}_{\mathrm{MDC}}}

\newcolumntype{C}[1]{>{\centering\arraybackslash$}p{#1}<{$}}

\def\squarebox#1{\hbox to #1{\hfill\vbox to #1{\vfill}}}

\newcommand{\qedblack}{\hspace*{\fill}%
  $\blacksquare$\smallskip}

\newcommand{\ZGibbs}{Z_{\mathrm{G}}}

\newtheorem{Lemma}{Lemma}
\newtheorem{Theorem}[Lemma]{Theorem}

\newtheorem{Example}[Lemma]{Example}

\newenvironment{Proof}%
  {\noindent \emph{Proof:}}{\qedblack}

\newcommand{\perm}{\operatorname{perm}}

\newcommand{\afwofp}[1]{a_{\mathrm{wofp}}}

\newcommand{\afbal}[1]{a_{\mathrm{bal}}}

\begin{document}
\title{Analysis of Double Covers of Factor Graphs}

\author{
  \IEEEauthorblockN{Pascal O. Vontobel}
  \IEEEauthorblockA{Department of Information Engineering \\
                    The Chinese University of Hong Kong \\
                    Shatin, N.T., Hong Kong \\
                    \texttt{pascal.vontobel@ieee.org}}
}

\maketitle

\watermark{\put(-30, 12){\versionstringone}
           \put(-30,  0){\versionstringtwo}
           \put(-30,-12){\versionstringthr}}

\begin{abstract}
  Many quantities of interest in communications, signal processing, artificial
  intelligence, and other areas can be expressed as the partition sum of some
  factor graph. Although the exact calculation of the partition sum is in many
  cases intractable, it can often be approximated rather well by the Bethe
  partition sum. In earlier work, we have shown that graph covers are a useful
  tool for expressing and analyzing the Bethe approximation. In this paper, we
  present a novel technique for analyzing double covers, a technique which
  ultimately leads to a deeper understanding of the Bethe approximation.
\end{abstract}

\section{Introduction}
\label{sec:introduction:1}

Consider a normal factor graph (NFG) $\graphN$~(see
\cite{Kschischang:Frey:Loeliger:01, Forney:01:1, Loeliger:04:1}).  Its
partition sum is defined to be
\begin{align}
  Z(\graphN)
    &\defeq
       \sum_{\va \in \setA}
         g(\va),
           \label{eq:partition:sum:1}
\end{align}
where the sum is over all configurations of $\graphN$ and where $g$ is the
global function of $\graphN$. For example, the constrained capacity of a
storage system can be expressed as the partition sum of a suitably formulated
NFG (see, \eg, \cite{Parvaresh:Vontobel:12:1, Vontobel:14:1}).

Because in many cases of interest the quantity $Z(\graphN)$ is intractable,
people have come up with various techniques for efficiently approximating
$Z(\graphN)$. For NFGs with non-negative-valued local functions, a popular
approach is to approximate $Z(\graphN)$ by the Bethe partition sum
$\ZBethe(\graphN)$, a quantity which is defined via the minimum of the Bethe
free energy function~\cite{Yedidia:Freeman:Weiss:05:1}. A reason for the
popularity of the Bethe approximation is that in many cases it can be found
efficiently with the help of the sum-product
algorithm~\cite{Kschischang:Frey:Loeliger:01, Loeliger:04:1,
  Yedidia:Freeman:Weiss:05:1}.

In contrast to the above, analytical definition of $\ZBethe(\graphN)$, it was
shown in~\cite{Vontobel:13:2} that $\ZBethe(\graphN)$ admits the following,
combinatorial characterization in terms of graph covers. Namely,
\begin{align}
  \ZBethe(\graphN)
    &= \limsup_{M \to \infty} \ 
         \ZBetheM{M}(\graphN),
           \label{eq:ZBethe:intro:1} \\
  \ZBetheM{M}(\graphN)
    &\defeq
       \sqrt[M]{\Big\langle \!
                  Z(\cgraph{N})
                \! \Big\rangle_{\cgraph{N} \in \cset{N}_{M}}}.
                             \label{eq:ZBethe:intro:2}
\end{align}
Here the expression under the root sign represents the (arithmetic) average of
$Z(\cgraph{N})$ over all $M$-covers $\cgraph{N}$ of $\graphN$, $M \geq 1$.

Note that we can write
\begin{align}
  \underbrace{
    \frac{Z(\graphN)}
         {\ZBethe(\graphN)}
  }_{\text{\ding{192}}}
    &= \underbrace{
         \frac{Z(\graphN)}
              {\ZBetheM{2}(\graphN)}
       }_{\text{\ding{193}}}
       \cdot
       \underbrace{
         \frac{\ZBetheM{2}(\graphN)}
              {\ZBethe(\graphN)}
       }_{\text{\ding{194}}} \ .
         \label{eq:Z:ratios:1}
\end{align}
For many NFGs, a significant contribution to the ratio~\ding{192} comes from
the ratio~\ding{193}. Therefore, understanding the ratio~\ding{193} can give
useful insights to understanding the ratio~\ding{192}.

The aim of the present paper is to develop techniques towards better
understanding and quantifying the ratio~\ding{193}. In particular, we will
study the partition sum of double covers of log-supermodular NFGs and thereby
give an alternative proof for a special case of a theorem by
Ruozzi~\cite{Ruozzi:12:1}.

On the one hand, the contributions here can be seen as adding another tool in
the holographic transformations toolbox for NFGs~\cite{AlBashabsheh:Mao:11:1,
  Forney:Vontobel:11:1}, and, on the other hand, they can be seen as adding
another tool to the toolbox for relating the partition sum and its Bethe
approximation~(see, \eg, \cite{Chertkov:Chernyak:06:1,
  Parvaresh:Vontobel:12:1, Vontobel:14:1, Mori:15:1}).

\subsection{Overview}
\label{sec:overview:1}

The paper is structured as follows. In
Section~\ref{sec:nfgs:and:their:covers:1} we give a brief introduction to NFGs
and their double covers. In Section~\ref{sec:analyzing:double:covers} we
present a novel technique for analyzing double covers. In
Section~\ref{sec:log:supermodular:nfgs:1} we apply this technique to the
analysis of a special class of log-supermodular NFGs. Finally, in
Section~\ref{sec:conclusion:and:outlook:1} we conclude the paper.

\section{Normal Factor Graphs and Their Finite Covers}
\label{sec:nfgs:and:their:covers:1}

Factor graphs are a convenient way to represent multivariate
functions~\cite{Kschischang:Frey:Loeliger:01}. In this paper we use a variant
called normal factor graphs (NFGs)~\cite{Forney:01:1} (also called
Forney-style factor graphs~\cite{Loeliger:04:1}), where variables are
associated with edges. The following example is taken
from~\cite{Vontobel:13:2}.

\begin{figure}
  \begin{center}
    \epsfig{file=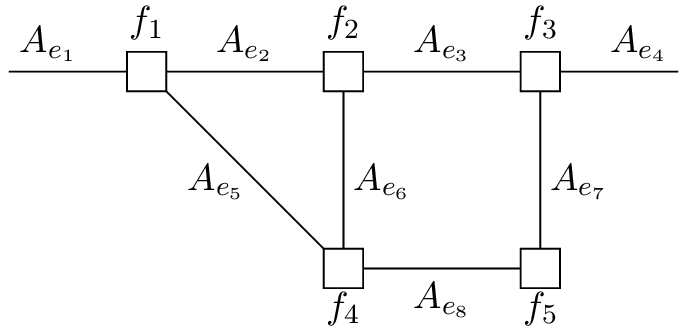, scale=0.88}
  \end{center}
  \caption{NFG $\graphN$ used in Example~\ref{example:simple:ffg:1}.}
  \label{fig:example:simple:ffg:1}
\end{figure}

\begin{Example}
  \label{example:simple:ffg:1}

  Consider the multivariate function
  \begin{align*}
    g(a_{e_1}, \ldots, a_{e_8})
      &\defeq
         f_1(a_{e_1}, a_{e_2}, a_{e_5})
         \cdot
         f_2(a_{e_2}, a_{e_3}, a_{e_6}) \, \cdot \\
      &\hspace{-1cm}
         f_3(a_{e_3}, a_{e_4}, a_{e_7})
         \cdot
         f_4(a_{e_5}, a_{e_6}, a_{e_8})
         \cdot
         f_5(a_{e_7}, a_{e_8}),
  \end{align*}
  where the so-called global function $g$ is the product of the so-called
  local functions $f_1$, $f_2$, $f_3$, $f_4$, and $f_5$. The decomposition of
  this global function as a product of local functions can be depicted with
  the help of an NFG $\graphN$ as shown in
  Fig.~\ref{fig:example:simple:ffg:1}. In particular, the NFG $\graphN$
  consists of
  \begin{itemize}
    
  \item the function nodes $f_1$, $f_2$, $f_3$, $f_4$, and $f_5$;

  \item the half edges $e_1$ and $e_4$ (sometimes also called ``external
    edges'');

  \item the full edges $e_2$, $e_3$, $e_5$, $e_6$, $e_7$, and $e_8$ (sometimes
    also called ``internal edges'').

  \end{itemize}
  In general,
  \begin{itemize}

  \item a function node $f$ represents the local function $f$;

  \item with an edge $e$ we associate the variable $A_e$ (note that a
    realization of the variable $A_e$ is denoted by $a_e$);

  \item an edge $e$ is incident on a function node $f$ if and only if $a_e$
    appears as an argument of the local function $f$.

  \end{itemize}
  Finally, we associate with $\graphN$ the partition sum~$Z(\graphN)$ as
  defined in~\eqref{eq:partition:sum:1}. (Note that we do not consider any
  temperature dependence of $Z(\graphN)$ in this paper.)
\end{Example}

\begin{figure}
  \begin{center}
    \epsfig{file=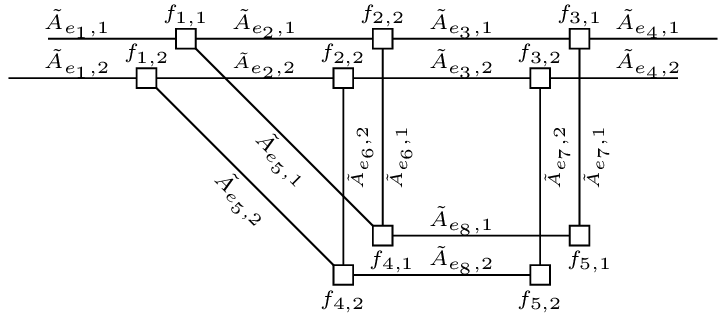, scale=0.88} \\[0.5cm]
    \epsfig{file=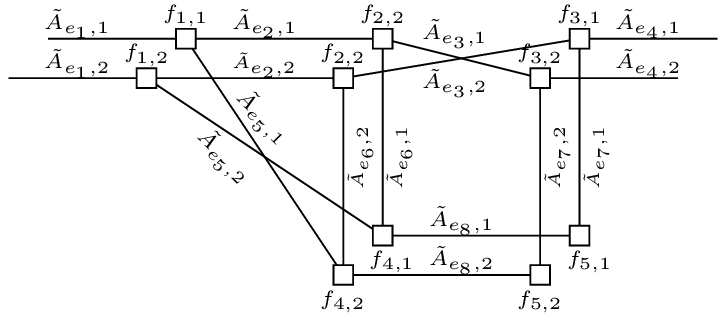, scale=0.88}
  \end{center}
  \caption{Two possible $2$-covers of the NFG $\graphN$ that is shown in
    Fig.~\ref{fig:example:simple:ffg:1}.}
  \label{fig:example:simple:ffg:1:cover:1}
\end{figure}

Throughout this paper, we will essentially use the same notation as
in~\cite{Vontobel:13:2}. The only exceptions are $Z(\graphN)$ instead of
$\ZGibbs(\graphN)$ for the partition sum, and $f$ instead of $g_f$ for local
functions. (For notations which are not defined in this paper, we refer the
reader to Sections~II and~IV of~\cite{Vontobel:13:2}.) Note that for the rest
of this paper, we assume that local functions in the base NFG $\graphN$ take
on only non-negative real values, \ie, $f(\va_f) \in \R_{\geq 0}$ for all $f$
and all $\va_f$.

Central to this paper are also finite graph covers of an NFG. (For the
definition of a finite graph cover, see, \eg,~\cite{Vontobel:13:2}.) The
following example is taken from~\cite{Vontobel:13:2}.

\begin{Example}
  \label{example:simple:ffg:1:cont:1}

  Consider again the NFG $\graphN$ that is discussed in
  Example~\ref{example:simple:ffg:1} and depicted in
  Fig.~\ref{fig:example:simple:ffg:1}. Two possible $2$-covers of this
  (base) NFG are shown in Fig.~\ref{fig:example:simple:ffg:1:cover:1}. The
  first graph cover is ``trivial'' in the sense that it consists of two
  disjoint copies of the NFG in Fig.~\ref{fig:example:simple:ffg:1}. The
  second graph cover is ``more interesting'' in the sense that the edge
  permutations are such that the two copies of the base NFG are
  intertwined. (Of course, both graph covers are equally valid.)
\end{Example}

Based on finite graph covers, one can define the degree-$M$ Bethe partition
sum $\ZBetheM{M}(\graphN)$ as in~\eqref{eq:ZBethe:intro:2} for any $M \geq
1$. With this, one can prove the alternative expression for $\ZBethe(\graphN)$
in~\eqref{eq:ZBethe:intro:1}. When considering the value of
$\ZBetheM{M}(\graphN)$ from $M = 1$ to $M = \infty$, one goes from
$Z(\graphN)$ to $\ZBethe(\graphN)$ as shown in
Fig.~\ref{fig:degree:M:Bethe:partition:function:1}.

\begin{figure}
  {\footnotesize
  \begin{alignat*}{2}
    &\big.
       \ZBetheM{M}(\graphN)
     \big|_{M \to \infty}
        &&= \ZBethe(\graphN) \\
    &\hskip1cm \Big\vert \\
    &\big.
       \ZBetheM{M}(\graphN)
     \big. && \\
    &\hskip1cm \Big\vert \\
    &\big.
       \ZBetheM{M}(\graphN)
     \big|_{M = 1}
       &&= Z(\graphN)
  \end{alignat*}
  }
  \caption{The degree-$M$ Bethe partition function of the NFG $\graphN$ for
    different values of $M$.}
  \label{fig:degree:M:Bethe:partition:function:1}
\end{figure}

In this paper, we also need the definition of a binary log-supermodular NFG:
it is an NFG with binary variables and log-supermodular local
functions. Recall that a local function $f: \{ 0, 1\}^{d_f} \to \R_{\geq 0}$
is called log-supermodular if
\begin{align*}
  f(\va'_f) \cdot f(\va''_f)
   &\leq 
      f(\va'_f \wedge \va''_f)
      \cdot
      f(\va'_f \vee \va''_f)
\end{align*}
holds for all $\va'_f, \va''_f \in \{ 0, 1 \}^{d_f}$, where
\begin{align*}
  (\va'_f \wedge \va''_f)_e 
    &\defeq 
       \min(a'_{f,e}, a''_{f,e}), 
         \quad e \in \set{E}_f, \\
  (\va'_f \vee \va''_f)_e 
    &\defeq 
       \max(a'_{f,e}, a''_{f,e}),
         \quad e \in \set{E}_f.
\end{align*}
Similarly, $f: \{ 0, 1\}^{d_f} \to \R_{\geq 0}$ is called log-submodular if
\begin{align*}
  f(\va'_f) \cdot f(\va''_f)
   &\geq
      f(\va'_f \wedge \va''_f)
      \cdot
      f(\va'_f \vee \va''_f)
\end{align*}
holds for all $\va'_f, \va''_f \in \{ 0, 1 \}^{d_f}$.

With a function like $f: \{ 0, 1 \}^2 \to \R$, it is natural to associate the
matrix
\begin{align*}
  \matr{T}_f
    &\defeq
       \begin{pmatrix}
         f(0,0) & f(0,1) \\
         f(1,0) & f(1,1)
       \end{pmatrix}.
\end{align*}
Note that the determinant of $\matr{T}_f$ is
\begin{align*}
  \det(\matr{T}_f) 
    &= f(0,0) \cdot f(1,1) 
       - 
       f(1,0) \cdot f(0,1).
\end{align*}
Clearly,
\begin{center}
  if $f$ is log-supermodular then $\det(\matr{T}_f) \geq 0$; \\
  if $f$ is log-submodular then $\det(\matr{T}_f) \leq 0$.
\end{center}

The following theorem was shown by Ruozzi~\cite{Ruozzi:12:1}. Its elegant
proof was based on the four-function theorem and generalizations thereof.

\begin{Theorem}[\!\!\cite{Ruozzi:12:1}]
  \label{theorem:log:supermodular:graph:cover:inequality:1}

  Let $\graphN$ be a binary log-supermodular NFG. Then for any $M$-cover
  $\cgraph{N}$ of $\graphN$, $M \geq 1$, it holds that
  \begin{align}
    Z(\cgraph{N})
      &\leq 
         Z(\graphN)^M.
           \label{eq:theorem:log:supermodular:graph:cover:inequality:1}
  \end{align}
\end{Theorem}

Combining~\eqref{eq:theorem:log:supermodular:graph:cover:inequality:1}
with~\eqref{eq:ZBethe:intro:2}, one obtains $\ZBetheM{M}(\graphN) \leq
Z(\graphN)$ for all $M \geq 1$. Moreover, using~\eqref{eq:ZBethe:intro:1}, one
obtains $\ZBethe(\graphN) \leq Z(\graphN)$. Note that before Ruozzi's paper,
the result $\ZBethe(\graphN) \leq Z(\graphN)$ had been proven by Sudderth
\etal~\cite{Sudderth:Wainwright:Willsky:07:1} for some special cases of binary
log-supermodular graphical models. After Ruozzi's paper, Weller and
Jebara~\cite{Weller:Jebara:14:1} gave an alternative proof for binary
log-supermodular NFGs where all function nodes (except the equality function
nodes) have degree two.

\section{Analyzing Double Covers}
\label{sec:analyzing:double:covers}

Consider an arbitrary NFG $\graphN$ without half edges,\footnote{Because we
  are mainly interested in the partition sum of $\graphN$ and because summing
  over variables associated with half edges is straightforward, considering
  only NFGs without half edges is no major restriction.} where $\set{A}_e
\defeq \{ 0, 1 \}$ for all edges $e \in \setE$. In this section we present a
novel approach for analyzing $Z(\cgraph{N})$ for some double cover
$\cgraph{N}$ of $\graphN$, ultimately towards comparing $\ZBetheM{2}(\graphN)$
with $Z(\graphN)$ and $\ZBethe(\graphN)$. This approach consists of two steps:
\begin{itemize}

\item In the first step, we associate a new NFG with $\cgraph{N}$. We will
  call it the merged double cover NFG (MDC-NFG) associated with $\cgraph{N}$
  and denote it by $\mdcnfg{N}$.

\item In the second step, we apply a suitable holographic
  transform~\cite{AlBashabsheh:Mao:11:1, Forney:Vontobel:11:1} to the
  MDC-NFG. The resulting NFG is called the transformed MDC-NFG and denoted by
  $\tmdcnfg{N}$. The key property of $\mdcnfg{N}$ and $\tmdcnfg{N}$ is
  \begin{align}
    Z(\cgraph{N})
      &= Z(\mdcnfg{N})
       = Z(\tmdcnfg{N}).
           \label{eq:dcnfg:tdcnfg:partition:sum:1}
  \end{align}

\end{itemize}

\begin{figure}
  \begin{center}
    \subfigure[Part of the base NFG $\graphN$.]
      {\epsfig{file=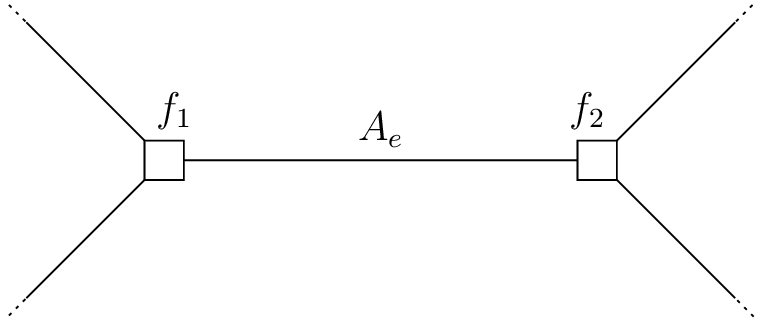, width=0.70\linewidth}}

    \subfigure[Part of possible double cover of $\graphN$.]
      {\epsfig{file=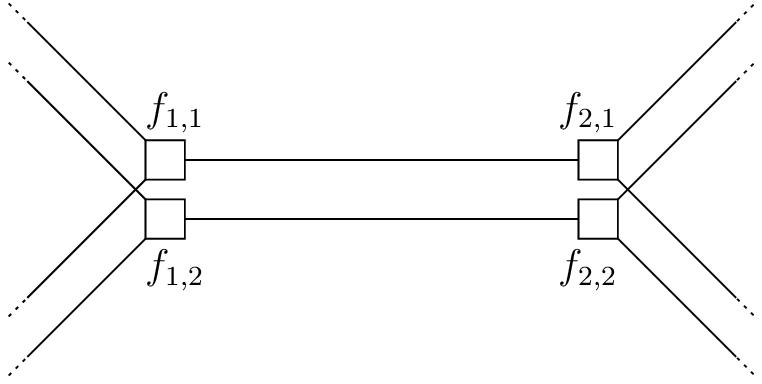, width=0.70\linewidth}}

    \subfigure[Part of possible double cover of $\graphN$.]
      {\epsfig{file=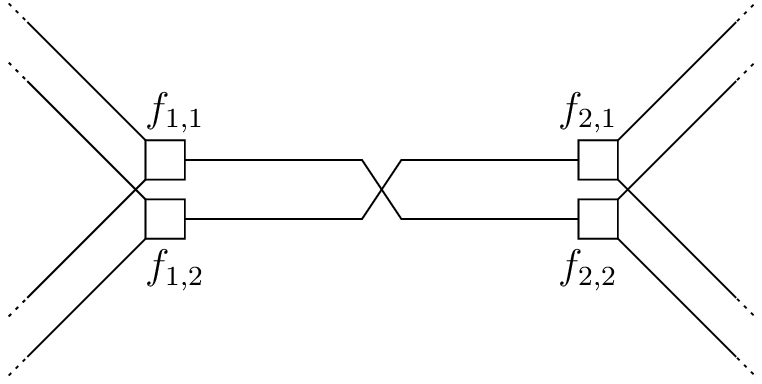, width=0.70\linewidth}}

    \subfigure[MDC-NFG.]
      {\epsfig{file=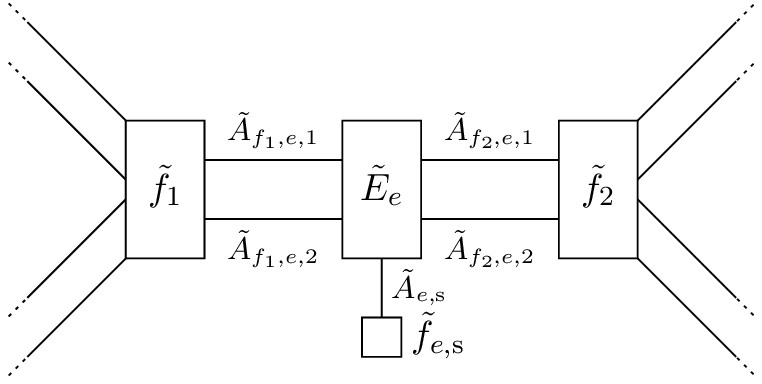, width=0.70\linewidth}}

    \subfigure[MDC-NFG with transform function nodes.]
      {\epsfig{file=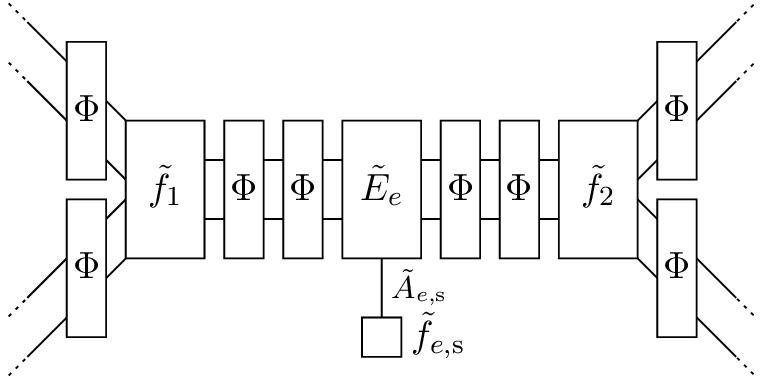, width=0.70\linewidth}}

    \subfigure[Transformed MDC-NFG.]
      {\epsfig{file=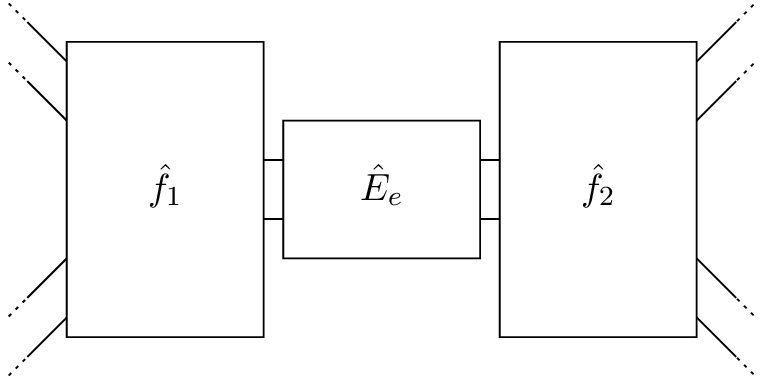, width=0.70\linewidth}}
  \end{center}
  \caption{Partial NFGs exemplifying the analysis technique in
    Section~\ref{sec:analyzing:double:covers}.}
  \label{fig:MDC:NFG:1}
\end{figure}

\noindent
The proposed approach is visualized in Fig.~\ref{fig:MDC:NFG:1} with the help
of an example NFG $\graphN$.
\begin{itemize}

\item Fig.~\ref{fig:MDC:NFG:1}(a) shows a part of a larger NFG
  $\graphN$. Here, $f_1$ and $f_2$ are function nodes of degree three.

\item Figs.~\ref{fig:MDC:NFG:1}(b) and~\ref{fig:MDC:NFG:1}(c) show the same
  part as in Fig.~\ref{fig:MDC:NFG:1}(a) for different double covers of
  $\graphN$.

\item Starting with a given double cover $\cgraph{N}$ of $\graphN$, the
  associated MDC-NFG $\mdcnfg{N}$ in Fig.~\ref{fig:MDC:NFG:1}(d) is obtained
  as follows.
  \begin{itemize}

  \item For every function node $f_j$ in $\graphN$ we close-the-box
    (see~\cite{Vontobel:Loeliger:02:2, Loeliger:04:1}) around every pair of
    function nodes $f_{j,1}$ and $f_{j,2}$ in $\cgraph{N}$ associated with
    $f_j$ and call the resulting function $\tilde f_j$. Because there are no
    variables to be summed over, $\tilde f_j$ is simply the product of
    $f_{j,1}$ and $f_{j,2}$. Note that if the function $f_j$ has $d_j$
    arguments, \ie, $f_j: \{ 0, 1 \}^{d_j} \to \R$, then
    \begin{align*}
      \tilde f_j: 
        \big\{ (0,0), \ (0,1), \ (1,0), \ (1,1) \big\}^{d_j} \to \R.
    \end{align*}

  \item For every edge $e$ in $\graphN$, we introduce the local function
    $\tilde E_e$ which encodes the non-crossing / the crossing of the pair of
    edges in $\cgraph{N}$ associated with $e$. The local function $\tilde E_e$
    is defined such that $\tilde a_{e,\mathrm{s}} = 0$ corresponds to the case
    where there is no crossing of the pair of edges in $\cgraph{N}$ and
    $\tilde a_{e,\mathrm{s}} = 1$ corresponds to the case where there is a
    crossing of the pair of edges in $\cgraph{N}$. With this, the matrices
    associated with
    \begin{align*}
      \tilde E_e
        \big( 
          (\tilde a_{f_1,e,1}, \tilde a_{f_1,e,2}), 
          (\tilde a_{f_2,e,1}, \tilde a_{f_2,e,2}), 
          \tilde a_{e,\mathrm{s}} \! = \! 0
        \big) \ , \\
      \tilde E_e
        \big( 
          (\tilde a_{f_1,e,1}, \tilde a_{f_1,e,2}),
          (\tilde a_{f_2,e,1}, \tilde a_{f_2,e,2}), 
          \tilde a_{e,\mathrm{s}} \! = \! 1
        \big) \phantom{ \ , }
    \end{align*}
    are, respectively,
    \begin{align*}
      \matr{\cover{E}}_{\mathrm{nocross}}
        &\defeq
           \left(
             \begin{smallmatrix}
               1 & 0 & 0 & 0 \\
               0 & 1 & 0 & 0 \\
               0 & 0 & 1 & 0 \\
               0 & 0 & 0 & 1
             \end{smallmatrix}
           \right),
      \quad
      \matr{\cover{E}}_{\mathrm{cross}}
         \defeq
           \left(
             \begin{smallmatrix}
               1 & 0 & 0 & 0 \\
               0 & 0 & 1 & 0 \\
               0 & 1 & 0 & 0 \\
               0 & 0 & 0 & 1
             \end{smallmatrix}
           \right).
    \end{align*}
    (The ``$\mathrm{s}$'' in $a_{e,\mathrm{s}}$ stands for ``switch.'')

  \item We define
    \begin{alignat*}{3}
      &
      \tilde f_{e,\mathrm{s}}(0) \defeq 1, \quad
      &&
      \tilde f_{e,\mathrm{s}}(1) \defeq 0
      &&
      \quad \text{(no crossing)}, \\
      &
      \tilde f_{e,\mathrm{s}}(0) \defeq 0, \quad
      &&
      \tilde f_{e,\mathrm{s}}(1) \defeq 1
      &&
      \quad \text{(crossing)}.
    \end{alignat*}

  \item One can verify that there is a bijection between valid configurations
    in $\cgraph{N}$ and valid configurations in $\mdcnfg{N}$, along with their
    corresponding global function values being equal. Therefore, $Z(\mdcnfg{N})
    = Z(\cgraph{N})$.


  \end{itemize}

\item The NFG in Fig.~\ref{fig:MDC:NFG:1}(f) is obtained from the NFG in
  Fig.~\ref{fig:MDC:NFG:1}(d) by introducing multiple instances of the
  function node $\Phi$ via opening-the-box. Here, the local function
  \begin{align*}
    \Phi: \big\{ (0,0), \ (0,1), \ (1,0), \ (1,1) \big\}^2 \to \R
  \end{align*}
  is specified via the matrix associated with $\Phi$, namely,
  \begin{align*}
    \matr{T}_{\Phi}
      &\defeq
         \left(
           \begin{array}{c|cc|c}
             1 & 0          & 0           & 0 \\
           \hline
           & & & \\[-0.35cm]
             0 & 1/\sqrt{2} & 1/\sqrt{2}  & 0 \\
             0 & 1/\sqrt{2} & -1/\sqrt{2} & 0 \\
           \hline
             0 & 0          & 0           & 1
           \end{array}
         \right).
  \end{align*}
  Note that $\matr{T}_{\Phi}^\tr = \matr{T}_{\Phi}$ and $\matr{T}_{\Phi}^{-1} =
  \matr{T}_{\Phi}$. 

\item Finally, the transformed MDC-NFG $\tmdcnfg{N}$ in
  Fig.~\ref{fig:MDC:NFG:1}(f) is obtained from Fig.~\ref{fig:MDC:NFG:1}(e) by
  applying several closing-the-box operations. Namely, for every edge $e \in
  E$, the function node $\hat{E}_{e}$ is obtained by closing-the-box around
  $\cover{E}_{e}$, the two adjacent $\Phi$-function nodes, and the $\tilde
  f_{e,\mathrm{s}}$ function node. With this, if the pair of edges in
  $\cgraph{N}$ corresponding to $e$ does not cross~/ does cross then the
  matrix associated with $\hat E_e$ equals, respectively,
  \begin{alignat*}{2}
    \matr{\hat{E}}_{\mathrm{nocross}}
      &\defeq
         \matr{T}_{\Phi}
           \cdot 
           \matr{\cover{E}}_{\mathrm{nocross}}
           \cdot 
           \matr{T}_{\Phi}
     &&= \left(
           \begin{array}{c|cc|c}
             1 & 0 & 0 & 0 \\
           \hline
             0 & 1 & 0 & 0 \\
             0 & 0 & 1 & 0 \\
           \hline
             0 & 0 & 0 & 1
           \end{array}
         \right), \\
    \matr{\hat{E}}_{\mathrm{cross}}
      &\defeq
    \matr{T}_{\Phi}
      \cdot 
      \matr{\cover{E}}_{\mathrm{cross}}
      \cdot 
      \matr{T}_{\Phi}
    &&= \left(
           \begin{array}{c|cc|c}
             1 & 0 &  0 & 0 \\
           \hline
             0 & 1 &  0 & 0 \\
             0 & 0 & -1 & 0 \\
           \hline
             0 & 0 &  0 & 1
           \end{array}
         \right).
  \end{alignat*}
  For every $f_j \in \set{F}$, the function node $\hat{f}_j$ is obtained by
  closing-the-box around the function node $\cover{f}_j$ and the $d_j$
  adjacent $\Phi$-function nodes, where $d_j$ is the degree of the function
  node $f_j$. The above construction implies (see~\cite{AlBashabsheh:Mao:11:1,
    Forney:Vontobel:11:1}) that $Z(\tmdcnfg{N}) = Z(\mdcnfg{N})$. Combining
  this with the equality $Z(\mdcnfg{N}) = Z(\cgraph{N})$, we
  obtain~\eqref{eq:dcnfg:tdcnfg:partition:sum:1}.

\end{itemize}

Let us conclude this section by considering a variation of the definition of
$\tilde f_{e,\mathrm{s}}$ and with that a variation of the definition of
$\tmdcnfg{N}$. Namely, for every $e \in \set{E}$, define $\tilde
f_{e,\mathrm{s}}(0) \defeq \frac{1}{2}$ and $\tilde f_{e,\mathrm{s}}(1) \defeq
\frac{1}{2}$. Then the matrix associated with $\hat E_e$ equals
\begin{align*}
  \matr{T}_{\hat{E}_e}
    &\defeq
       \frac{1}{2}
         \cdot
         \matr{\hat{E}}_{\mathrm{nocross}}
       +
       \frac{1}{2}
         \cdot
         \matr{\hat{E}}_{\mathrm{cross}}
     = \left(
           \begin{array}{c|cc|c}
             1 & 0 &  0 & 0 \\
           \hline
             0 & 1 &  0 & 0 \\
             0 & 0 &  0 & 0 \\
           \hline
             0 & 0 &  0 & 1
           \end{array}
         \right).
\end{align*}
These definitions allow one to formulate the following theorem, whose proof we
omit.

\begin{Theorem}
  For the NFG $\tmdcnfg{N}$ as just specified, it holds that
  \begin{align}
    \ZBetheM{2}(\graphN)
      &= \sqrt{Z(\tmdcnfg{\graphN})} \ .
           \label{eq:theorem:atdcnfg:1}
  \end{align}
\end{Theorem}

Note that, in contrast to~\eqref{eq:ZBethe:intro:2}, only a single NFG appears
in the expression on the right-hand side of~\eqref{eq:theorem:atdcnfg:1}.

\section{Log-Supermodular NFGs}
\label{sec:log:supermodular:nfgs:1}

In this section we apply the technique from
Section~\ref{sec:analyzing:double:covers} to analyze the following class of
NFGs: it consists of all NFGs without half edges and
\begin{itemize}

\item where $\set{A}_e = \{ 0, 1 \}$ for all edges $e \in \set{E}$,

\item where all local functions are log-supermodular, and

\item where all function nodes have degree $2$ or $3$, except for equality
  indicator function nodes which may have arbitrary degree at least
  $2$.\footnote{Note that equality indicator functions are log-supermodular.}

\end{itemize}

\begin{Theorem}
  \label{theorem:log:supermodular:NFG:double:cover:inequality:1}

  Let $\graph{N}$ be an NFG from the class of NFGs specified above let
  $\cgraph{N}$ be an arbitrary double cover of $\graphN$. Then
  \begin{align*}
    Z(\cgraph{N})
      &\leq
         Z(\graphN)^2.
  \end{align*}
\end{Theorem}

\begin{Proof}
  (sketch) Let $\tmdcnfg{N}$ and $\tmdcnfg{N}^{\mathrm{trivial}}$ be
  associated with the double cover $\cgraph{N}$ and the trivial double cover
  $\cgraph{N}^{\mathrm{trivial}}$, respectively. Their partition sums are
  $Z(\tmdcnfg{N}) = \sum_{\hva} \hat g(\hva)$ and
  $Z(\tmdcnfg{N}^{\mathrm{trivial}}) = \sum_{\hva} \hat
  g^{\mathrm{trivial}}(\hva)$, respectively. Note that both sums are over the
  same set of configurations. From the results in
  Sections~\ref{sec:analyzing:double:covers} and the upcoming results in
  Section~\ref{sec:log:supermodular:nfgs:1} it follows that $\hat
  g^{\mathrm{trivial}}(\hva) \geq 0$ and $\hat g(\hva) = \pm \hat
  g^{\mathrm{trivial}}(\hva)$ for all $\hva$, which implies $Z(\tmdcnfg{N})
  \leq Z(\tmdcnfg{N}^{\mathrm{trivial}})$. Finally, because $Z(\tmdcnfg{N}) =
  Z(\cgraph{N})$ and $Z(\tmdcnfg{N}^{\mathrm{trivial}}) =
  Z(\cgraph{N}^{\mathrm{trivial}}) = Z(\graphN)^2$, we obtain the promised
  result.
\end{Proof}

\subsection{Arbitrary Log-Supermodular Function Node of Degree $2$}

Let $f$ be a log-supermodular function with two arguments; let $t_{00} \defeq
f(0,0)$, $t_{01} \defeq f(0,1)$, $t_{10} \defeq f(1,0)$, $t_{11} \defeq
f(1,1)$. With this, the matrices associated with $f$, $\cover{f}$, and
$\hat{f}$ are, respectively,
\begin{align*}
  \matr{T}_f
    &\defeq
       \begin{pmatrix}
         t_{00} & t_{01} \\
         t_{10} & t_{11}
       \end{pmatrix} \! ,
  \ 
  \matr{T}_{\cover{f}}
    \defeq
       \begin{pmatrix}
         t_{00} t_{00} & t_{00} t_{01} & t_{01} t_{00} & t_{01} t_{01} \\
         t_{00} t_{10} & t_{00} t_{11} & t_{01} t_{10} & t_{01} t_{11} \\
         t_{10} t_{00} & t_{10} t_{01} & t_{11} t_{00} & t_{11} t_{01} \\
         t_{10} t_{10} & t_{10} t_{11} & t_{11} t_{10} & t_{11} t_{11}
       \end{pmatrix} \! ,
\end{align*}
{\scriptsize
\begin{align*}
  \matr{T}_{\hat{f}}
    &\defeq
       \matr{T}_{\Phi}
         \cdot 
         \matr{T}_{\cover{f}} 
         \cdot 
         \matr{T}_{\Phi}
     = \left(
         \begin{array}{c|cc|c}
           t_{00} t_{00} 
             & \sqrt{2} \, t_{00} t_{01} 
             & 0 
             & t_{01} t_{01} \\[0.08cm]
         \hline
         & & & \\[-0.25cm]
           \sqrt{2} \, t_{00} t_{10} 
             & \perm(\matr{T}_f) 
             & 0 
             & \sqrt{2} \, t_{01} t_{11} \\
           0 
             & 0 
             & \det(\matr{T}_f) 
             & 0 \\[0.08cm]
         \hline
         & & & \\[-0.25cm]
           t_{10} t_{10} 
             & \sqrt{2} \, t_{10} t_{11} 
             & 0 
             & t_{11} t_{11}
         \end{array}
       \right) \! ,
\end{align*}
}

\noindent
where $\perm(\matr{T}_f) \defeq t_{00} t_{11} + t_{10} t_{01}$. Because $f$ is
log-supermodular, $\det(\matr{T}_f)$ is non-negative, and so all entries of
$\matr{T}_{\hat{f}}$ are non-negative.

\subsection{Arbitrary Log-Supermodular Function Node of Degree $3$}

Let $f(a_1,a_2,a_3)$ be a log-supermodular function with three arguments and
let $t_{000} \defeq f(0,0,0)$, $t_{001} \defeq f(0,0,1)$, \etc\ Moreover, let
$\matr{T}_{f|a_3=0}$ and $\matr{T}_{f|a_3=1}$ be the matrices associated with
the functions $f(a_1,a_2,0)$ and $f(a_1,a_2,1)$, respectively. (Clearly, if
$f(a_1,a_2,a_3)$ is a log-supermodular function, then also $f(a_1,a_2,0)$ and
$f(a_1,a_2,1)$ are log-supermodular functions.) The matrices
$\matr{T}_{f|a_1=0}$, $\matr{T}_{f|a_1=1}$, $\matr{T}_{f|a_2=0}$, and
$\matr{T}_{f|a_2=1}$ are defined analogously. Then the $4 \times 4 \times 4$
array $\matr{T}_{\hat{f}}$ associated with $\hat{f}$ is given by

{\scriptsize

\begin{center}
  $\displaystyle
  \left(
    \begin{array}{C{1.70cm}|C{1.70cm}C{1.70cm}|C{1.70cm}}
      t_{000} t_{000} 
        & \sqrt{2} \, t_{000} t_{010}
        & 0
        & t_{010} t_{010} \\[0.08cm]
    \hline
     & & & \\[-0.25cm]
      \sqrt{2} \, t_{000} t_{100} 
        & \perm(\matr{T}_{f|a_3=0})
        & 0 
        & \sqrt{2} \, t_{010} t_{110} \\
      0 
        & 0 
        & \det(\matr{T}_{f|a_3=0}) 
        & 0 \\[0.08cm]
    \hline
    & & & \\[-0.25cm]
      t_{100} t_{100} 
        & \sqrt{2} \, t_{100} t_{110} 
        & 0 
        & t_{110} t_{110}
    \end{array}
  \right)$, \\[0.35cm]
  $\displaystyle
  \left(
    \begin{array}{C{1.70cm}|C{1.70cm}C{1.70cm}|C{1.70cm}}
      \sqrt{2} \, t_{000} t_{001} 
        & \perm(\matr{T}_{f|a_1=0}) 
        & 0 
        & \sqrt{2} \, t_{010} t_{011} \\[0.08cm]
    \hline
    & & & \\[-0.25cm]
      \perm(\matr{T}_{f|a_2=0}) 
        & \hcf\big( \hzero, \hzero, \hzero \big)
        & 0 
        & \perm(\matr{T}_{f|a_2=1})  \\
      0 
        & 0 
        & \hcf\big( \hone, \hone, \hzero \big)
        & 0 \\[0.08cm]
    \hline
    & & & \\[-0.25cm]
      \sqrt{2} \, t_{100} t_{101} 
        & \perm(\matr{T}_{f|a_1=1}) 
        & 0 
        & \sqrt{2} \, t_{110} t_{111}
    \end{array}
  \right)$, \\[0.35cm]
  $\displaystyle
  \left(
    \begin{array}{C{1.70cm}|C{1.70cm}C{1.70cm}|C{1.70cm}}
      0 
        & 0 
        & \det(\matr{T}_{f|a_1=0}) 
        & 0 \\[0.08cm]
    \hline
    & & & \\[-0.25cm]
      0 
        & 0 
        & \hcf\big( \hzero, \hone, \hone \big)
        & 0 \\
      \det(\matr{T}_{f|a_1=0}) 
        & \hcf\big( \hone, \hzero, \hone \big)
        & 0 
        & \det(\matr{T}_{f|a_1=0}) \\[0.08cm]
    \hline
    & & & \\[-0.25cm]
      0 
        & 0 
        & \det(\matr{T}_{f|a_1=1}) 
        & 0
    \end{array}
  \right)$, \\[0.35cm]
  $\displaystyle
  \left(
    \begin{array}{C{1.70cm}|C{1.70cm}C{1.70cm}|C{1.70cm}}
      t_{001} t_{001} 
        & \sqrt{2} \, t_{001} t_{011} 
        & 0 
        & t_{011} t_{011} \\[0.08cm]
    \hline
    & & & \\[-0.25cm]
      \sqrt{2} \, t_{001} t_{101} 
        & \perm(\matr{T}_{f|a_3=1}) 
        & 0 
        & \sqrt{2} \, t_{011} t_{111} \\
      0 
        & 0 
        & \det(\matr{T}_{f|a_3=1}) 
        & 0 \\[0.08cm]
    \hline
    & & & \\[-0.25cm]
      t_{101} t_{101} 
        & \sqrt{2} \, t_{101} t_{111} 
        & 0 
        & t_{111} t_{111}
    \end{array}
  \right)$, \\[0.35cm]
\end{center}

}

\noindent
where
\begin{align*}
  \hcf(\hzero,\hzero,\hzero)
    &= \gamma
       \cdot
       (
         t_{000} t_{111} 
         + 
         t_{100} t_{011} 
         +
         t_{010} t_{101}
         +
         t_{000} t_{110}
       ), \\
  \hcf(\hone,\hzero,\hone)
    &= \gamma
       \cdot
       (
         t_{000} t_{111} 
         -
         t_{100} t_{011} 
         +
         t_{010} t_{101}
         -
         t_{001} t_{110}
       ), \\
  \hcf(\hzero,\hone,\hone)
    &= \gamma
       \cdot
       (
         t_{000} t_{111} 
         +
         t_{100} t_{011} 
         -
         t_{010} t_{101}
         -
         t_{001} t_{110}
       ), \\
  \hcf(\hone,\hone,\hzero)
    &= \gamma
       \cdot
       (
         t_{000} t_{111} 
         - 
         t_{100} t_{011} 
         -
         t_{010} t_{101}
         +
         t_{001} t_{110}
       ),
\end{align*}
and where $\hzero \defeq (0,1)$, $\hone \defeq (1,0)$, and $\gamma \defeq 1 /
\sqrt{2}$.

\begin{Lemma}
  All entries of $\matr{T}_{\hat{f}}$ are non-negative.
\end{Lemma}

\begin{Proof}
  For most entries of $\matr{T}_{\hat{f}}$ the statement is clearly
  true. Moreover, the log-supermodularity of $f$ implies that all entries
  based on determinants must be non-negative. Also, from the definition of
  $\hcf(\hzero,\hzero,\hzero)$, it follows that $\hcf(\hzero,\hzero,\hzero)
  \geq 0$. It only remains to show $\hcf(\hone, \hzero,\hone) \geq 0$,
  $\hcf(\hzero,\hone, \hone) \geq 0$, and $\hcf(\hone, \hone, \hzero) \geq
  0$. In this proof we show $\hcf(\hzero,\hone,\hone) \geq 0$. Analogous lines
  of reasoning yield $\hcf(\hone,\hzero,\hone) \geq 0$ and
  $\hcf(\hone,\hone,\hzero) \geq 0$.
  
  Let $s_0 \defeq \gamma \cdot t_{000} t_{111}$, $s_1 \defeq \gamma \cdot
  t_{100} t_{011}$, $s_2 \defeq \gamma \cdot t_{010} t_{101}$, $s_3 \defeq
  \gamma \cdot t_{001} t_{110}$. From log-supermodularity of $f$ it follows
  that $s_0 \geq s_1$, \ $s_0 \geq s_2$, \ $s_0 \geq s_3$, and $s_0 s_1 \geq
  s_2 s_3$.

  The inequality $\hcf(\hzero,\hone,\hone) \geq 0$ is equivalent to the
  inequality $s_0 + s_1 - s_2 - s_3 \geq 0$. We show the latter inequality by
  considering two cases: $0 \! \leq \! s_2 \! \leq \! s_1 \! \leq \! s_0$ and
  $0 \! \leq \! s_1 \! < \! s_2 \! \leq \! s_0$.
  \begin{itemize}

  \item Assume $0 \leq s_2 \leq s_1 \leq s_0$. Then $s_0 + s_1 - s_2 - s_3
    \geq 0$ follows immediately from the combination of $s_0 \geq s_3$ and
    $s_1 \geq s_2$.
  
  \item Assume $0 \leq s_1 < s_2 \leq s_0$. Then $(s_0 - s_2) \cdot (s_2 -
    s_1) \geq 0$ implies $s_0 s_2 + s_1 s_2 - s_2^2 - s_0 s_1 \geq 0$. Using
    $s_0 s_1 \geq s_2 s_3$, this inequality implies $s_0 s_2 + s_1 s_2 - s_2^2
    - s_2 s_3 \geq 0$, which in turn implies $s_0 + s_1 - s_2 - s_3 \geq 0$
    because $s_2 > 0$.
  
  \end{itemize}
\end{Proof}

\subsection{Equal Function Node of Arbitrary Degree At Least $2$}

We have the following theorem, whose proof is omitted.

\begin{Theorem}

  Let $f$ be an equality indicator function with $d \geq 2$ arguments. Then
  $\hat f \big( (\hat a_{1,1}, \hat a_{1,2}), \ldots, (\hat a_{d,1}, \hat
  a_{d,2}) \big)$ equals
  {\footnotesize
  \begin{align*}
      &  \left\{
           \begin{array}{ll}
           1 & \text{if $\hat a_{i,m} = 0 \ \forall \ i \in \{ 1, \ldots, d \},
                                                   \ m \in \{ 1, 2 \}$} \\
           1 & \text{if $\hat a_{i,m} = 1 \ \forall \ i \in \{ 1, \ldots, d \},
                                                   \ m \in \{ 1, 2 \}$} \\
           2^{1-d/2}
             & \text{if $(\hat a_{i,1}, \hat a_{i,2}) 
                       \in \big\{ (0,1), \ (1,0) \big\} \ 
                       \forall \ i \in \{ 1, \ldots, d \}$} \\
             & \text{and $\sum_{i=1}^{d} \hat a_{i,1} = 0 
                     \ (\mathrm{mod} \ 2)$} \\
           0 & \text{otherwise}
           \end{array}
         \right.
  \end{align*}  
  }  
\end{Theorem}

\section{Conclusion and Outlook}
\label{sec:conclusion:and:outlook:1}

We leave it as an open problem to generalize
Theorem~\ref{theorem:log:supermodular:NFG:double:cover:inequality:1} to all
binary log-supermodular NFGs, \ie, to the setup of
Theorem~\ref{theorem:log:supermodular:graph:cover:inequality:1}. Moreover, we
will discuss elsewhere how the results in
Sections~\ref{sec:analyzing:double:covers}
and~\ref{sec:log:supermodular:nfgs:1} can be used to quantify the
ratio~\ding{193} in~\eqref{eq:Z:ratios:1}.

\section*{Acknowledgment}
\label{sec:ack:1}

It is a pleasure to acknowledge discussions on the topic of this paper with
Chun Lam Chan, Mahdi Jafari, Sid Jaggi, and Navin Kashyap.

\bibliographystyle{IEEEtran}
\bibliography{/home/vontobel/references/references}

\end{document}